\documentclass[aps, pra, twocolumn, amsmath, amssymb, showpacs, 
]{revtex4-1}

\usepackage[utf8]{inputenc}
\usepackage{graphicx}
\usepackage{units}
\usepackage{color}
\usepackage[pdftex, colorlinks=true, linkcolor=myblue, citecolor=myblue,
urlcolor=myblue]{hyperref}
\usepackage{times}
\usepackage{bbold}
\usepackage{comment}

\definecolor{myblue}{rgb}{0,0,1}

\begin{document}

\title{Decay of dark and bright plasmonic modes in a metallic nanoparticle dimer}

\author{Adam Brandstetter-Kunc}
\affiliation{Institut de Physique et Chimie des Mat\'{e}riaux de Strasbourg, Universit\'{e} de
Strasbourg, CNRS UMR 7504, F-67034 Strasbourg, France}

\author{Guillaume Weick}
\email{guillaume.weick@ipcms.unistra.fr}
\affiliation{Institut de Physique et Chimie des Mat\'{e}riaux de Strasbourg, Universit\'{e} de
Strasbourg, CNRS UMR 7504, F-67034 Strasbourg, France}

\author{Dietmar Weinmann}
\affiliation{Institut de Physique et Chimie des Mat\'{e}riaux de Strasbourg, Universit\'{e} de
Strasbourg, CNRS UMR 7504, F-67034 Strasbourg, France}

\author{Rodolfo A.\ Jalabert}
\affiliation{Institut de Physique et Chimie des Mat\'{e}riaux de Strasbourg, Universit\'{e} de
Strasbourg, CNRS UMR 7504, F-67034 Strasbourg, France}


\begin{abstract}
We develop a general quantum theory of the coupled plasmonic modes resulting from the 
near-field interaction between localized surface plasmons in a heterogeneous metallic nanoparticle
dimer. In particular, we provide analytical expressions for the frequencies and
decay rates of the bright and dark plasmonic modes.
We show that, for sufficiently small nanoparticles, the main decay
channel for the dark plasmonic mode, which is weakly coupled to light and, hence, immune to radiation damping, is of
nonradiative origin and corresponds to Landau damping,
i.e., decay into electron-hole pairs.
\end{abstract}


\maketitle

\section{Introduction}
It has taken 74 years between the founding work of Mie on the optical response of a metallic
nanoparticle \cite{mie08_AP} and the extension of Ruppin to the case of two nearby spheres 
\cite{ruppi82_PRB}. 
At the practical level, the evolution from the single object to compound optical resonant systems has
been even slower than the corresponding theoretical development. About
twenty centuries span from the realization of optically active materials based on 
noninteracting nanoparticles \cite{frees07} to the success in the fabrication and optical measurements of
ensembles of interacting nanoparticles \cite{kreibig}. 
Nonetheless, once the theoretical and experimental basis for studying these compound objects was
laid down, the subsequent developments have been extremely fast. In the field of nanoplasmonics
\cite{maier}, the intense 
recent activity concerning nanoparticle dimers \cite{jain10_CPL} stems from the fact
that it is the simplest system sustaining coupled plasmonic excitations.

The near-field interaction between the localized surface plasmons (LSPs) of two nanoparticles
results in a bright mode (coupled to the electromagnetic field associated with visible light) and a
dark one (weakly coupled to light). Both of these modes have been experimentally
observed \cite{tamar02_APL, rechb03_OC, olk08_NL, jain10_CPL, chu09_NL, koh09_ACS, barrow14_NL} and theoretically investigated \cite{ruppi82_PRB, gerar83_PRB, nordl04_NL,
dahme07_NL, bache08_PRL, zuloa09_NL, esteb12_NatureComm, zhang14_preprint}. On the one hand, 
the bright mode has been observed using laser excitation in various experimental
systems \cite{tamar02_APL, rechb03_OC, olk08_NL, jain10_CPL}. 
On the other hand, the dark mode is difficult to excite in symmetric, homogeneous dimers with interparticle
distance much smaller than the laser wavelength. However, this difficulty is less severe in heterogeneous
dimers. The alternative experimental technique of electron energy loss spectroscopy (EELS) has
recently provided an unambiguous detection of the dark mode
\cite{chu09_NL, koh09_ACS, barrow14_NL}. 

The damping of these coupled modes is a crucial limiting factor for their experimental 
observation as well
as for potential applications in the field of nanoplasmonics \cite{maier}. 
While the bright mode radiates in the
far-field and hence has a radiative decay, the dark mode is obviously immune to radiation damping.
It is then of paramount interest to understand the nonradiative decay channels at the origin of the experimentally-observed finite
linewidth of the dark mode \cite{chu09_NL, koh09_ACS, barrow14_NL}. 

In this work we show that for dimers composed of sufficiently small nanoparticles, the main decay channel for the dark mode 
corresponds to Landau damping, 
which dominates over absorption losses. 
In the present context, the Landau damping is a purely quantum-mechanical effect that leads to the
decay of the collective excitation through the creation of electron-hole pairs \cite{kawab66_JPSJ, bertsch}.
Thus, we
develop a general quantum theory of coupled plasmonic excitations in a heterogeneous dimer of
metallic nanoparticles. Using bosonic Bogoliubov transformations and semiclassical techniques, we
provide analytical expressions for the frequencies and lifetimes of the coupled plasmonic modes. 

The present paper is organized as follows: Section \ref{sec:model} presents our model that we use in
Secs.\ \ref{sec:freq} and \ref{sec:decay} to obtain the frequencies and the decay rates of the
coupled plasmonic modes, respectively. We draw our conclusions in Sec.\ \ref{sec:ccl}. The technical
details of our calculations are presented in the appendices.

\section{Open quantum system approach}
\label{sec:model}
For a single metallic nanoparticle, the separation of the electronic
coordinates into center-of-mass and relative motion \cite{gerch02_PRA,
weick06_PRB} amounts to a description
typical for an open quantum system. The dipolar LSP (i.e., the
center-of-mass coordinate) is coupled to an electronic environment (i.e., the
bath of electron-hole pairs represented by the relative coordinates) and leads
to the nonradiative decay of the collective excitation (Landau damping). The coupling between the two
subsystems  is a consequence of the breaking of Kohn's theorem \cite{kohn61_PR,
jacak} due to the
non-harmonicity of the confining potential, the latter arising from the positive
ionic background. In addition, radiative damping arises from the coupling of the
LSP with electromagnetic field modes, while absorption (Ohmic) losses occur due to the finite resistivity of
the metal.

\begin{figure}[tb]
\includegraphics[width=\columnwidth]{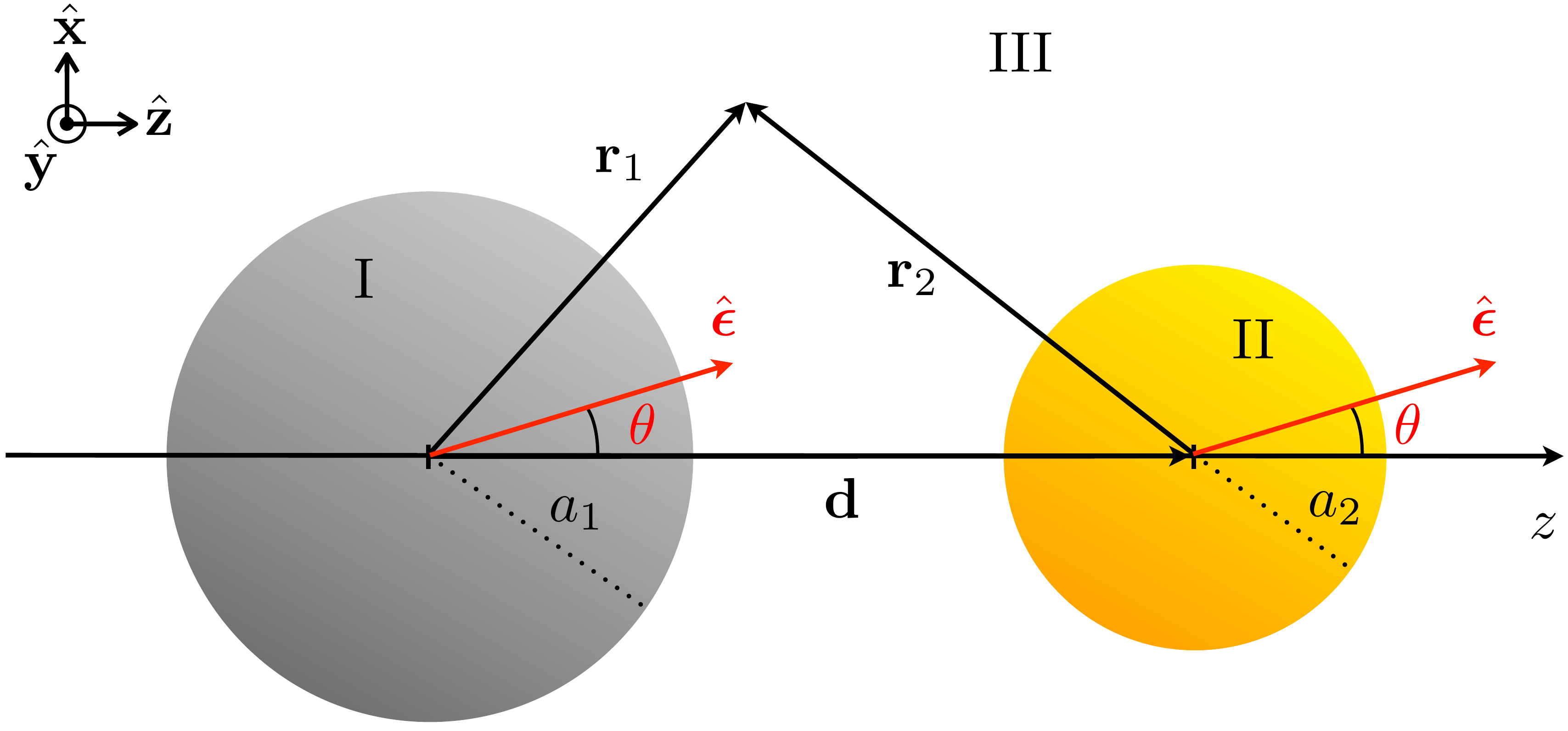}
\caption{\label{fig:NPs}%
(Color online).
Sketch of a nanoparticle dimer formed by two spherical nanoparticles of radii
$a_1$ and $a_2$ separated by a distance $d$, together with the coordinate system 
used in the text. The polarization $\hat{\boldsymbol{\epsilon}}$ of the localized 
surface plasmons forming an angle $\theta$ with the $z$ axis is also shown.}
\end{figure}

As detailed in Appendix \ref{sec:micro}, extending this approach to the case of a nanoparticle dimer (sketched in Fig.\ \ref{fig:NPs}), the resulting electronic 
Hamiltonian can be written as 
\begin{equation}
\label{eq:H_decompo}
H=H_\mathrm{pl}+H_\mathrm{eh}+H_\mathrm{pl-eh}.
\end{equation}
The plasmonic  part reads
\begin{equation}
\label{eq:H_pl}
H_\mathrm{pl}=\sum_{n=1}^2\hbar\tilde{\omega}_n b_n^\dagger b_n^{\phantom{\dagger}}
+\hbar\Omega f(\theta)\left(b_1^\dagger b_2^{\phantom{\dagger}}+b_1^\dagger
b_2^\dagger+\mathrm{h.c.}\right),
\end{equation}
where the index $n$ is used to identify within the dimer the two spherical, neutral
nanoparticles of radius $a_n$ (each containing $N_n$ electrons).
The LSP frequency
\begin{equation}
\label{eq:omega_Mie}
\tilde{\omega}_n=\omega_{n}\sqrt{1-\frac{N_{\mathrm{out},n}}{N_n}},\quad
\omega_{n}=\sqrt{\frac{3N_ne^2}{m_\mathrm{e}a_n^3\left(\epsilon_\mathrm{d}^{(n)}+2\epsilon_\mathrm{m}\right)}}
\end{equation}
 is redshifted
with respect to the Mie frequency
$\omega_{n}$
due to the $N_{\mathrm{out},n}$ electrons spilling out of nanoparticle $n$
\cite{kreibig}. Here, $-e$ and $m_\mathrm{e}$ denote the electron charge and mass,
respectively. The dielectric constant $\epsilon_\mathrm{d}^{(n)}$ takes into
account the screening provided, in the case of noble metals,
by the d electrons in nanoparticle $n$, and
$\epsilon_\mathrm{m}$ is the dielectric constant of the matrix in which the nanoparticles are embedded. 
In Eq.\ \eqref{eq:H_pl}, the bosonic operator $b_n$ ($b_n^\dagger$) annihilates (creates)
an LSP in nanoparticle $n$ \cite{bergm03_PRL}. The two LSPs interact through their near fields,
giving rise to the second term in the right-hand side of Eq.\ \eqref{eq:H_pl},
where
\begin{equation}
\label{eq:Omega}
\Omega=\frac
12\prod_{n=1}^2\left(\frac{\tilde{\omega}_n}{1-N_{\mathrm{out},n}/N_n}\right)^{1/2}
\left(\frac{a_n}{d}\right)^{3/2}
\end{equation}
and 
\begin{equation}
f(\theta)=1-3\cos^2{\theta}. 
\end{equation}
Here, $d$ is the center-to-center nanoparticle
distance and $\theta$ is the angle formed by the polarization
$\hat{\boldsymbol{\epsilon}}$ of the LSPs and the $z$ axis joining the two NPs (see Fig.\
\ref{fig:NPs}). In writing Eq.\ \eqref{eq:H_pl}, we adopted a quasistatic
dipole-dipole approximation valid for $3a_n\lesssim d\ll c/\tilde\omega_n$,
where $c$ is the speed of light \cite{brong00_PRB, park04_PRB}. 
We further assumed that in each eigenmode, the 
two LSPs are polarized in the same direction $\hat{\boldsymbol{\epsilon}}$.

Electron-hole excitations within each nanoparticle provide the electronic
environment described by \cite{weick06_PRB}
\begin{equation}
\label{eq:H_env}
H_\mathrm{eh}=\sum_{n=1}^2\sum_\alpha\varepsilon_{n\alpha}^{} c_{n\alpha}^\dagger
c_{n\alpha}^{\phantom{\dagger}},
\end{equation}
where $c_{n\alpha}^{}$ ($c_{n\alpha}^\dagger$) annihilates (creates) an electron
in the $n$th nanoparticle
associated with the one-body state $|n\alpha\rangle$ with energy
$\varepsilon_{n\alpha}$ in the self-consistent potential $V$. 
Note that the form \eqref{eq:H_env} implicitly assumes that tunneling of
electrons between the two nanoparticles is suppressed. 
Similarly to the case of a single nanoparticle discussed above, the
coupling of the plasmon to the electronic environment comes from the non-harmonicity of the
single-particle confinement, which in the jellium approximation with
$\epsilon_\mathrm{d}^{(n)}=\epsilon_\mathrm{m}=1$ reads
\begin{equation}
\label{eq:U}
U_n(r_n)=\frac{N_ne^2}{2a_n^3}(r_n^2-3a_n^2)\Theta(a_n-r_n)-\frac{N_ne^2}{r_n}\Theta(r_n-a_n),
\end{equation}
where $r_n$ is the radial coordinate with respect to the center of nanoparticle $n$.
Hence, the Hamiltonian $H_\mathrm{pl-eh}$ in Eq.\
\eqref{eq:H_decompo} can be written as 
\begin{align}
\label{eq:H_c}
H_\mathrm{pl-eh}=&\sum_{n, n', n''=1}^2
\sqrt{\frac{\hbar}{2N_nm_\mathrm{e}\tilde\omega_n}}
\left(b_n^{\phantom{\dagger}}+b_n^\dagger\right)
\nonumber\\
&\times
\sum_{\alpha\beta}
\langle n'\alpha|\hat{\boldsymbol{\epsilon}}\cdot\nabla U_n(r_n)|n'' \beta\rangle
c_{n'\alpha}^\dagger c_{n''\beta}^{\phantom{\dagger}}.
\end{align}

\section{Frequencies of the coupled plasmonic modes}
\label{sec:freq}

\begin{figure}[tb]
\includegraphics[width=\columnwidth]{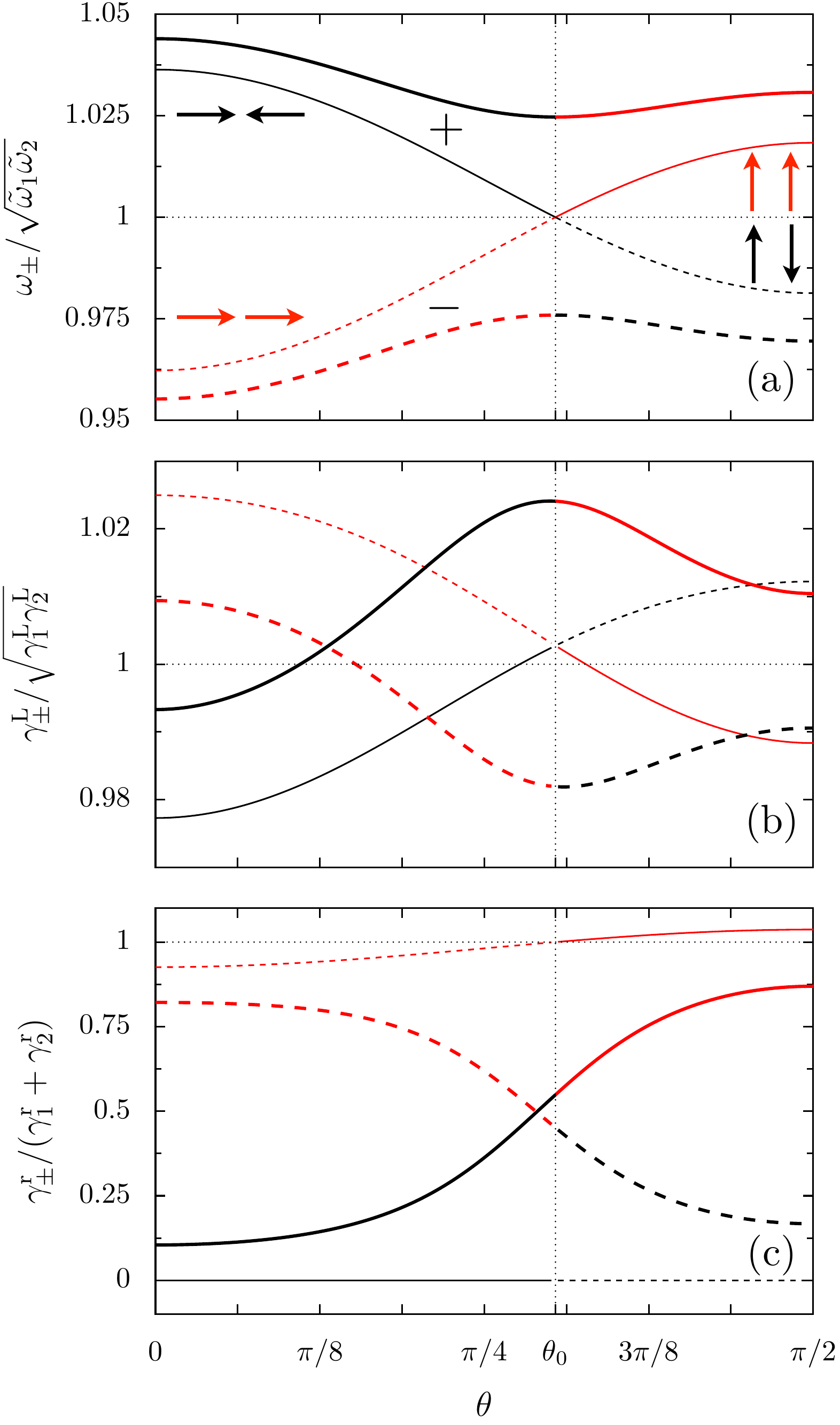}
\caption{\label{fig:omega_pm}%
(Color online). 
(a) Frequencies $\omega_\pm$ [Eq.\ \eqref{eq:omega_pm}], (b) Landau damping linewidths $\gamma_\pm^\mathrm{L}$ [Eq.\ \eqref{eq:gamma_pm_L}] and (c)
radiative damping linewidths $\gamma_\pm^\mathrm{r}$ [Eq.\ \eqref{eq:gamma_pm_r}]
of the $+$ (solid lines) and $-$ (dashed lines) coupled plasmonic modes as a function of the polarization
angle $\theta$ for $\tilde\omega_1/\tilde\omega_2=1$ (thin lines) and
$\tilde\omega_1/\tilde\omega_2=1.05$ (thick lines). The bright (dark) modes for which the two LSPs
are in phase (in anti-phase) are represented by red/gray (black) curves.
In the figure, the parameters are $a_1=a_2=a$, $d=3a$,
$\hbar\tilde\omega_1/E_\mathrm{F}^{(1)}=1$, and spill-out is neglected.}
\end{figure}

The quadratic Hamiltonian \eqref{eq:H_pl} representing the two coupled LSPs is diagonalized as 
\begin{equation}
H_\mathrm{pl}=\sum_{\sigma=\pm}\hbar\omega_\sigma B_\sigma^\dagger
B_\sigma^{\phantom{\dagger}}
\end{equation}
 by 
introducing the bosonic operators
\begin{equation}
\label{eq:capital_B}
B_\pm=\sum_{n=1}^2(u_{n,\pm} b_n+\bar u_{n,\pm} b_n^\dagger).
\end{equation}
For the general case of unequal frequencies $\tilde\omega_n$, 
following Tsallis' prescription for Bogoliubov transformations \cite{tsall78_JMP}, we find (see Appendix \ref{sec:diag} for details)
\begin{equation}
\label{eq:omega_pm}
\omega_\pm=\sqrt{\frac{\tilde{\omega}_1^2+\tilde{\omega}_2^2}{2}
\pm\sqrt{4\Omega^2\tilde{\omega}_1\tilde{\omega}_2f^2(\theta)
+\left(\frac{\tilde{\omega}_1^2-\tilde{\omega}_2^2}{2}\right)^2}} 
\end{equation}
and 
\begin{subequations}
\label{eq:coeff}
\begin{align}
u_{n,\pm}=[\pm\ \mathrm{sign}\{f(\theta)\}]^{n-1}
\frac{\omega_\pm+\tilde{\omega}_n}{2\sqrt{\tilde{\omega}_n\omega_\pm}}
\sqrt{\frac{\omega_\pm^2-\tilde{\omega}_{\hat n}^2}{2\omega_\pm^2-\tilde{\omega}_1^2-\tilde{\omega}_2^2}},
\\
\bar u_{n,\pm}=[\pm\ \mathrm{sign}\{f(\theta)\}]^{n-1}
\frac{\omega_\pm-\tilde{\omega}_n}{2\sqrt{\tilde{\omega}_n\omega_\pm}}
\sqrt{\frac{\omega_\pm^2-\tilde{\omega}_{\hat n}^2}{2\omega_\pm^2-\tilde{\omega}_1^2-\tilde{\omega}_2^2}}.
\end{align}
\end{subequations}
In Eq.\ \eqref{eq:coeff}, $\hat n=1(2)$ for $n=2(1)$. 

The two plasmonic eigenmodes correspond to the coherent oscillation of the two
LSPs. For $\theta=0$, the low-energy (high-energy) mode with frequency $\omega_-$ ($\omega_+$) can be thought of as the
in-phase (anti-phase) motion of the two LSPs. Vice versa, for $\theta=\pi/2$,
the $-$ and $+$ modes correspond to the anti-phase and in-phase motions,
respectively. Figure \ref{fig:omega_pm}(a) shows the transition between these two
previous extreme cases as a function of the polarization angle $\theta$. In the
special case $\tilde\omega_1=\tilde\omega_2$ [i.e., identical nanoparticles,
thin solid and dashed lines in Fig.\ \ref{fig:omega_pm}(a)],
the in-phase mode (with nonvanishing dipole moment) can be excited by dipolar
light and thus receives the name of ``bright mode". It corresponds to the
$-$ ($+$) eigenmode for polarization angles $\theta<(>)\theta_0$, where 
$\theta_0=\arccos{(1/\sqrt{3})}$ is the angle for which the dipole-dipole
interaction in Eq.\ \eqref{eq:H_pl} vanishes. Conversely, the anti-phase mode
(with vanishing dipole moment) corresponds to the $+$ ($-$) eigenmode for
$\theta<(>)\theta_0$. Since it cannot be triggered by visible light, it is
referred to as the ``dark mode".
When $\tilde\omega_1\neq\tilde\omega_2$ [thick, solid and dashed
lines in Fig.\ \ref{fig:omega_pm}(a)], the difference between bright and dark
modes is less stringent, as both the $+$ and $-$ modes have a finite dipole
moment for any $\theta$. In this case the usage
of bright (dark) modes refers to the larger (smaller) total dipole moment. 
Notice, moreover, that the dependence on the interparticle distance $d$ of the
$\pm$ frequencies is encapsulated in Eq.\ \eqref{eq:omega_pm} in the definition \eqref{eq:Omega} of
$\Omega$, so that $\omega_\pm-[(\tilde\omega_1^2+\tilde\omega_2^2)/2]^{1/2}\sim
\pm 1/d^3$ \cite{footnote:d}. Such a behavior, which directly follows from the form of the dipole-dipole
interaction, has also been unveiled both theoretically \cite{ruppi82_PRB,
nordl04_NL} and experimentally \cite{rechb03_OC} in the case of nanoparticles
of equal size and formed of the same material.

\section{Nonradiative and radiative decay rates of the dark and bright modes}
\label{sec:decay}

\subsection{Landau damping}
The modes previously described can be understood as resulting from the coupling
of classical dipoles, as has been extensively discussed in the literature
\cite{bache08_PRL, chu09_NL, dahme07_NL, ruppi82_PRB, gerar83_PRB, jain10_CPL,
koh09_ACS, nordl04_NL, olk08_NL, rechb03_OC, tamar02_APL, kreibig, barrow14_NL}.
Our quantum description is nevertheless crucial for the evaluation of the
Landau damping of the two coupled plasmonic modes.
The coupling Hamiltonian \eqref{eq:H_c} associated with this decay channel can be expressed in terms
of the $B_\pm$ bosonic operators given in Eq.\ \eqref{eq:capital_B} as
\begin{equation}
\label{eq:H_pl-eh}
H_\mathrm{pl-eh}=
\sum_{n=1}^2\sum_{\sigma=\pm}\Lambda_n\left(B_\sigma^{\phantom{\dagger}}+B_\sigma^\dagger\right)
\sum_{\alpha\beta}\hat{\boldsymbol{\epsilon}}\cdot\mathbf{D}_{\alpha\beta,\sigma}^{(n)}
c_{n\alpha}^\dagger c_{n\beta}^{}, 
\end{equation}
with
\begin{align}
\label{eq:D}
\mathbf{D}_{\alpha\beta,\sigma}^{(n)}=&\;
\Delta u_{n,\sigma}\mathbf{d}_{\alpha\beta}^{(n)}
+\frac{2\Omega}{\omega_{n}}
\Delta u_{\hat n,\sigma}\left[\mathbf{d}_{\alpha\beta}^{(n)}
-3\left(\mathbf{d}_{\alpha\beta}^{(n)}\cdot\hat{\mathbf{z}}\right)\hat{\mathbf{z}}\right], 
\end{align}
where $\Lambda_n=(\hbar m_\mathrm{e}\omega_n^3/2N_n)^{1/2}$ and $\Delta u_{n,\sigma}=u_{n,\sigma}-\bar u_{n,\sigma}$.
Equation \eqref{eq:H_pl-eh} is obtained under the assumption that the self-consistent potential $V$
is constant inside the nanoparticles and infinite outside. Such an assumption, which neglects the
spill-out, is justified by
density functional calculations for the one-particle case \cite{weick06_PRB}, as well as for dimers \cite{zuloa09_NL}. 
Within this approximation, the dipole matrix elements entering Eq.\ \eqref{eq:D} reads 
\begin{equation}
\label{eq:d}
\mathbf{d}_{\alpha\beta}^{(n)}=\left(
\sum_{s=\pm1}\mathcal{A}_{l_\alpha l_\beta, s}^{m_\alpha m_\beta}
\frac{\hat{\mathbf{x}}-\mathrm{i}s\hat{\mathbf{y}}}{\sqrt{2}}
+\mathcal{A}_{l_\alpha l_\beta, 0}^{m_\alpha m_\beta}\hat{\mathbf{z}}\right)
\mathcal{R}_n(E_\alpha,E_\beta),
\end{equation}
where 
the radial part is given by \cite{yanno92_AP}
\begin{equation}
\mathcal{R}_n(E_\alpha, E_\beta)=\frac{2\hbar^2}{m_\mathrm{e}a_n}
\frac{\sqrt{E_\alpha E_\beta}}{(E_\alpha-E_\beta)^2}.
\end{equation}
The angular part in Eq.\ \eqref{eq:d} is expressed in terms of Wigner-$3j$ symbols as \cite{weick11_PRB}
\begin{align}
\label{eq:angular}
\mathcal{A}_{l_\alpha l_\beta,s}^{m_\alpha m_\beta}=&\;
(-1)^{m_\alpha+s}\sqrt{(2l_\alpha+1)(2l_\beta+1)}
\nonumber\\
&\times
\begin{pmatrix}
l_\alpha & l_\beta & 1 \\
0 & 0 & 0
\end{pmatrix}
\begin{pmatrix}
l_\alpha & l_\beta & 1 \\
-m_\alpha & m_\beta & s
\end{pmatrix}.
\end{align}
Notice that the angular momenta selection rules $l_\alpha=l_\beta\pm1$ and $m_\alpha=m_\beta$
($s=0$) and $m_\alpha=m_\beta\pm1$ ($s=\pm1$) are encapsulated in the expression above. 

The zero-temperature Fermi's golden rule decay rate of the $+$ and $-$ plasmonic modes from the
Landau damping channel is then given from Eq.\ \eqref{eq:H_pl-eh} by 
\begin{equation}
\label{eq:FGR}
\gamma_\pm^\mathrm{L}=\frac{2\pi}{\hbar}\sum_{n=1}^2\sum_{eh}
|\Lambda_n\mathbf{D}_{eh,\pm}^{(n)}\cdot\hat{\boldsymbol{\epsilon}}|^2
\delta(\hbar\omega_\pm-E_e+E_h), 
\end{equation}
where $|n e\rangle$ and $|n h\rangle$ represent, respectively, electron and hole states in the
self-consistent potential $V$ for the $n$th nanoparticle. The sum over $e$ and $h$ states is performed by introducing the density of states
$\varrho_l^{(n)}(E)$ with fixed angular momentum $l$ at energy $E$ in nanoparticle $n$. The angular momentum selection
rules from Eq.\ \eqref{eq:angular} lead to 
\begin{align}
\label{eq:gamma_inter}
\gamma_\pm^\mathrm{L}=&\;
\frac{16\pi}{3\hbar m_\mathrm{e}^2\omega_\pm^4}\sum_{n=1}^2\frac{\Lambda_n^2}{a_n^2}\mathcal{P}_{n,\pm}(\theta)
\int_{\max{\{E_\mathrm{F}^{(n)},\hbar\omega_\pm\}}}^{E_{\mathrm{F}}^{(n)}+\hbar\omega_\pm}
\mathrm{d}E\, EE_\pm
\nonumber\\
&\times
\sum_l\varrho_l^{(n)}(E)
\left[(l+1)\varrho^{(n)}_{l+1}(E_\pm)+l\varrho^{(n)}_{l-1}(E_\pm)
\right],
\end{align}
with $E_\pm=E-\hbar\omega_\pm$ and 
where 
\begin{align}
\mathcal{P}_{n,\pm}(\theta)=&\;
\sin^2{\theta}
\left(
\Delta u_{n,\pm}+\frac{2\Omega}{\omega_n}\Delta u_{\hat n,\pm}
\right)^2
\nonumber\\
&+\cos^2{\theta}
\left(
\Delta u_{n,\pm}-\frac{4\Omega}{\omega_n}\Delta u_{\hat n,\pm}
\right)^2.
\end{align}
Here, $E_\mathrm{F}^{(n)}$ ($v_\mathrm{F}^{(n)}$) stands for the Fermi energy (velocity) in nanoparticle
$n$.
Using the semiclassical leading-order form \cite{weick05_PRB} of the density of states,
\begin{equation}
\varrho_l^{(n)}(E)\simeq
\frac{\sqrt{2m_\mathrm{e}a_n^2E/\hbar^2-(l+1/2)^2}}{2\pi E},
\end{equation}
the Landau damping decay rates read 
\begin{equation}
\label{eq:gamma_pm_L}
\gamma_\pm^\mathrm{L}=
\sum_{n=1}^2\frac{3v_\mathrm{F}^{(n)}}{4a_n}\left(\frac{\omega_n}{\omega_\pm}\right)^3
g\left({\hbar\omega_\pm}/{E_\mathrm{F}^{(n)}}\right)\mathcal{P}_{n,\pm}(\theta), 
\end{equation}
where 
an explicit expression of the function
\begin{equation}
g(\nu)=\frac{2}{\nu}\int_{\max\{1, \nu\}}^{1+\nu}\mathrm{d}x
\int_0^{x-\nu}\mathrm{d}y \sqrt{(x-y)(x-y-\nu)}
\end{equation}
can be found in Refs.\ \cite{yanno92_AP, weick11_PRB},
thus yielding an analytical expression for the
Landau damping decay rates. 

The linewidths from Eq.\ \eqref{eq:gamma_pm_L} are represented as a function of
the polarization angle 
$\theta$ in Fig.\ \ref{fig:omega_pm}(b) for the case
$\tilde\omega_1/\tilde\omega_2=1$ (thin lines) and
$\tilde\omega_1/\tilde\omega_2=1.05$ (thick lines). The dark (black) and bright
modes (red/gray lines) show a modulation with respect to the Landau damping
linewidth of isolated nanoparticles \cite{kawab66_JPSJ, yanno92_AP, weick05_PRB}
\begin{equation}
\label{eq:kubo}
\gamma_n^\mathrm{L}=\frac{3v_\mathrm{F}^{(n)}}{4a_n}g\left(\hbar\omega_n/E_\mathrm{F}^{(n)}\right),
\end{equation}
used as normalization. 
This anisotropy represents a qualitative difference as compared to the single nanoparticle case,
stemming from the nonlocality of the coupled plasmonic modes. The expected tunability of
$\pm\unit[2]{\%}$ obtained for $d=3a$ [Fig.\ \ref{fig:omega_pm}(b)] should be detectable in optical
experiments for the bright mode.
When $\tilde\omega_1/\tilde\omega_2=1$, the higher
energy $+$ mode is less damped than the lower-energy $-$ one. This energy
dependence is analogous to the single nanoparticle case, where higher mode frequencies correspond to
lower values of the damping rates \cite{seoan07_EPJD, footnote:theta_0}.

\subsection{Absorption losses and radiation damping}
In order to assess the relevance of Landau damping, we have to quantify the additional
damping mechanisms not described by the Hamiltonian \eqref{eq:H_decompo}. The absorption
losses given by the bulk conductivity of the metal lead to a size-independent decay rate
$\gamma^\mathrm{a}$ (which has a weak frequency dependence). The radiation damping rate
$\gamma_\pm^\mathrm{r}$ relates the power $P^\mathrm{r}_\pm$ radiated with 
the energy $E_\pm$ stored in the mode $\pm$ as 
\begin{equation}
\label{eq:power_radiated}
P^\mathrm{r}_\pm=\gamma_\pm^\mathrm{r}E_\pm^{}.
\end{equation}
In the limit where the interparticle distance is much smaller than the wavelength associated
with each LSP \cite{smith10_EPJ}, one has 
$P^\mathrm{r}_\pm=2\omega_\pm^4p_\pm^2/3c^3$, where $p_\pm$ is the dipole
moment corresponding to the $\pm$ mode oscillating at the frequency $\omega_\pm$
given by Eq.\ \eqref{eq:omega_pm}.
Averaging Eq.\ \eqref{eq:power_radiated} on a period $2\pi/\omega_\pm$ much shorter
than the decay time $1/\gamma_\pm^\mathrm{r}$ leads to 
\begin{equation}
\label{eq:gamma_pm_r}
\gamma_\pm^\mathrm{r}=\frac{2\omega_\pm^{3}}{3c^3}\left(\sum_{n=1}^2\sqrt{\tilde\omega_n
a_n^3}\ \Delta u_{n,\pm}\right)^2.
\end{equation}
The radiation damping linewidths above are shown in Fig.\ \ref{fig:omega_pm}(c)
as a function of polarization for $\tilde\omega_1/\tilde\omega_2=1$ (thin lines)
and $\tilde\omega_1/\tilde\omega_2=1.05$ (thick lines). The normalization factor $\gamma_1^\mathrm{r}+\gamma_2^\mathrm{r}$ used
in the figure  corresponds to the
radiation damping of two independent nanoparticles, with
\begin{equation}
\gamma_n^\mathrm{r}=\frac{2\tilde\omega_n^4a_n^3}{3c^3}. 
\end{equation}
For
$\tilde\omega_1/\tilde\omega_2=1$, the dark mode has a vanishing radiative
linewidth [thin black curve in Fig.\ \ref{fig:omega_pm}(c)] as it does not
couple to the electromagnetic field, while the radiation damping of the bright
mode can be modulated with the light polarization (thin red/gray curve). Choosing
$\tilde\omega_1/\tilde\omega_2\neq1$ opens the way to the optical detection of the
dark mode, as its linewidth becomes finite [thick black curve in Fig.\
\ref{fig:omega_pm}(c)].

\subsection{Discussion}
\begin{figure}[tb]
\includegraphics[width=\columnwidth]{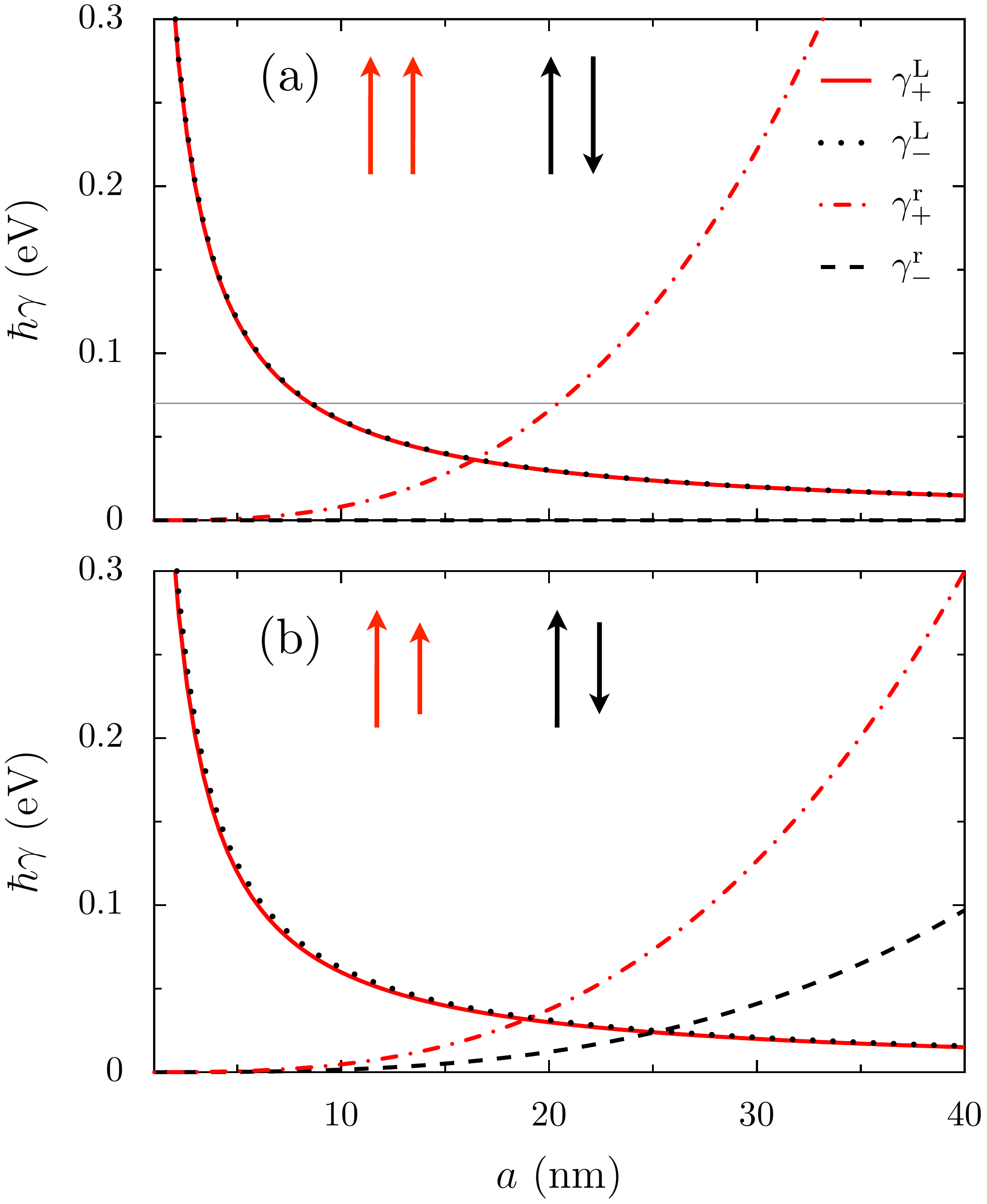}
\caption{\label{fig:gamma}%
(Color online). 
Landau damping (solid and dotted lines) and radiation damping (dashed-dotted and dashed lines) linewidths for transverse 
polarization $\theta=\pi/2$ as a function of nanoparticle radius $a$ of the bright ($+$, light
gray/red lines) and dark ($-$, black lines) mode.
(a) Homogeneous dimer composed of two Ag nanoparticles. The thin gray line in the figure corresponds to the absorption losses measured in Ref.\ \cite{charl89_ZPD}. (b) Heterogeneous Ag-Au dimer.
In the figure, $d=3a$ and $\epsilon_\mathrm{m}=4$.}
\end{figure}

The relative importance of the three above-described damping mechanisms, as well as those arising
from the nature of the embedding matrix (chemical interface damping and conduction band in the
matrix \cite{kreibig}) depends on a variety of physical parameters that should be settled in order
to achieve a meaningful comparison. We focus from now on on noble metal
nanoparticles \cite{footnote:noble}, since they
constitute the dimers experimentally studied \cite{tamar02_APL, rechb03_OC, olk08_NL, jain10_CPL, chu09_NL, koh09_ACS, barrow14_NL}. 
We show in Fig.\ \ref{fig:gamma} the competition of 
$\gamma_\pm^\mathrm{L}$ and $\gamma_\pm^\mathrm{r}$
of the bright (light gray/red lines) and dark (black lines)
plasmonic modes  
as a function of nanoparticle radius $a$ (assumed to be the same
for both particles) for homogeneous [Ag-Ag, Fig.\ \ref{fig:gamma}(a)] and heterogeneous [Ag-Au,
Fig.\ \ref{fig:gamma}(b)] dimers with $d=3a$. A matrix with $\epsilon_\mathrm{m}=4$ is assumed,
which leads to LSP resonances $\omega_\mathrm{Ag}=\unit[2.6]{eV/\hbar}$ and
$\omega_\mathrm{Au}=\unit[2.2]{eV/\hbar}$ \cite{kreibig}.
We consider the transverse polarization ($\theta=\pi/2$), so that the $+$ ($-$) mode corresponds
to the bright (dark) mode (see Fig.\ \ref{fig:omega_pm}). 
As can be seen from Fig.\ \ref{fig:gamma}(a), for the bright mode the Landau damping [Eq.\ \eqref{eq:gamma_pm_L} scaling with 
the nanoparticle size as $1/a$, solid line] dominates over radiation damping [Eq.\ \eqref{eq:gamma_pm_r} scaling as
$a^3$, dashed-dotted line] for $a\lesssim\unit[15]{nm}$.
On the contrary, for the dark mode, Landau
damping [dotted line in Fig.\ \ref{fig:gamma}(a)] dominates for sizes up to which it becomes 
negligible as compared to the absorption
losses, since the
radiative contribution to the linewidth vanishes [see dashed line in Fig.\ \ref{fig:gamma}(a)].
In the case of a heterogeneous dimer [Fig.\ \ref{fig:gamma}(b)], the difference
between the bright and dark plasmonic modes is less stringent, as the dark mode
acquires a finite dipole moment due to the 
difference in sizes and/or in densities between the two nanoparticles. 

The agreement of our analytical theory with microscopic numerical calculations \cite{zuloa09_NL} is
excellent. Using the time-dependent local-density approximation for Ag dimers with
$a=\unit[1.2]{nm}$, $d=3a$ and $\theta=0$, a resonance linewidth of $\unit[0.43]{eV}$ is obtained, while
the Landau damping mechanism, dominating in this regime, yields [Eq.\ \eqref{eq:gamma_pm_L}]
$\hbar\gamma_-^\mathrm{L}=\unit[0.40]{eV}$. 

The existing experimental data exhibit tendencies that are consistent with our theoretical
calculations. In Ag dimers excited by EELS \cite{koh09_ACS} the bright and dark modes have both an
increasing damping rate when passing from homogeneous to heterogeneous dimers, due to the larger
dipole moments of the latter and the fact that the inhomogeneous dimers are achieved by using larger
nanoparticles. Homogeneous dimers with $a=\unit[12]{nm}$ have a larger damping rate for the bright
mode than for the dark one, due to the radiation damping contribution on the former. 
However, a quantitative comparison of the damping rates is handicapped by the
limited resolution ($\sim\unit[0.2]{eV}$) of these EELS experiments  \cite{kriva14_Nature}.
Another difficulty for the quantitative comparison with the experiment is that the employed 
nanoparticles are very close to each other,
taking the setup outside the validity of the dipole-dipole approximation used in our theoretical approach.
Moreover, the absorption losses, estimated \cite{charl89_ZPD} to be about 
$\hbar\gamma^\mathrm{a}\simeq\unit[70]{meV}$ in optically excited Ag nanoparticles [see the thin gray line in Fig.\ \ref{fig:gamma}(a)],
are expected to be considerably larger in EELS experiments. This is due to the strong heating induced by the electron beam that might explain the value of the 
observed \cite{koh09_ACS} total linewidths ($\sim \unit[0.5]{eV}$).
In addition, the nature and dielectric properties of the material
coating the nanoparticles are not well controlled.

\section{Conclusion}
\label{sec:ccl}
We have presented a general quantum theory of coupled plasmonic
modes in a heterogeneous metallic nanoparticle dimer. We have
provided analytical expressions for the frequencies, Landau damping and
radiative linewidths of these plasmonic modes. 
The role of nonradiative damping for collective excitations of interacting
metallic nanoparticles has been explored and quantified. 
In particular, we have shown that the Landau damping
is an unavoidable decay channel for the dark plasmonic mode
consistent with the tendencies of the experimentally-observed linewidths \cite{chu09_NL,
koh09_ACS, barrow14_NL}. 
Our work should motivate systematic measurements for different particle sizes and constitutes a 
first step of crucial importance towards the
understanding of the damping mechanisms limiting plasmon propagation in
technologically promising quantum metamaterials based on one- and
two-dimensional arrays of nanoparticles \cite{maier, meinz14_NatPhoton}, such as the honeycomb lattice
supporting chiral massless Dirac-like plasmons~\cite{weick13_PRL, sturg14_preprint}.

\begin{acknowledgments}
We are grateful to J.\ A.\ Bad\'an, W.\ L.\ Barnes, E.\ Mariani, R.\ E.\
Marotti, and F.\ Vall\'ee for helpful discussions. We acknowledge
financial support from Centre National de la Recherche Scientifique (CNRS)
through the PICS program (Contract No.\ 6384 APAG) and from
Agence Nationale de la Recherche (ANR) under grant No.\ ANR-14-CE26-0005
Q-MetaMat.
\end{acknowledgments}

\appendix
\section{Microscopic Hamiltonian of a metallic nanoparticle dimer}
\label{sec:micro}

In this appendix, we detail the derivation of the Hamiltonian \eqref{eq:H_decompo}, which describes the bright and dark plasmonic modes and their 
coupling to electron-hole excitations. 
Within the jellium model which replaces the ions by a homogeneous positively charged background, 
the electronic Hamiltonian describing the nanoparticle dimer sketched in Fig.\ \ref{fig:NPs} reads
\begin{align}
\label{eq:H_original}
H=&
\sum_{n=1}^2\sum_{i=1}^{N_n}\left[\frac{\mathbf{p}_{n,i}^2}{2m_\mathrm{e}}+U_\mathrm{2NP}(\boldsymbol{\rho}_{n,i})\right]
\nonumber\\
&+\frac{e^2}{2}\sum_{n=1}^2\sum_{\substack{i,j=1\\(j\neq i)}}^{N_n}
\frac{1}{|\boldsymbol{\rho}_{n,i}-\boldsymbol{\rho}_{n,j}|}
+e^2\sum_{i=1}^{N_1}\sum_{j=2}^{N_2}
\frac{1}{|\boldsymbol{\rho}_{1,i}-\boldsymbol{\rho}_{2,j}|}
, 
\end{align}
with $\boldsymbol{\rho}_{n,i}$ the position of the $i$th electron belonging to the
$n$th nanoparticle and $\mathbf{p}_{n,i}$ its momentum.
Note that the Hamiltonian \eqref{eq:H_original} describes a nanoparticle dimer in vacuum
($\epsilon_\mathrm{m}=1$) in which the screening due to the d electrons is negligible
($\epsilon_\mathrm{d}^{(n)}=1$).
The third and fourth terms in the right-hand side in Eq.\ \eqref{eq:H_original}
represent, respectively, the intra- and interparticle electron-electron
interaction.
The single-particle confinement potential created by the two
positively-charged jellium
spheres (with charge $+N_ne$, $n=1,2$) reads
\begin{equation}
\label{eq:U_2NP}
U_\mathrm{2NP}(\boldsymbol{\rho})=
\begin{cases}
\displaystyle
\frac{N_1e^2}{2a_1} \left[\left(\frac{r_1}{a_1}\right)^2-3\right]-\frac{N_2 e^2}{r_2}, &
\boldsymbol{\rho}\in\textrm{I}, 
\vspace{.2truecm}
\\
\displaystyle
\frac{N_2e^2}{2a_2} \left[\left(\frac{r_2}{a_2}\right)^2-3\right]-\frac{N_1 e^2}{r_1}, &
\boldsymbol{\rho}\in\textrm{II}, 
\vspace{.2truecm}
\\
\displaystyle
-\frac{N_1e^2}{r_1}-\frac{N_2e^2}{r_2},&
\boldsymbol{\rho}\in\textrm{III}, 
\end{cases}
\end{equation}
where $\mathbf{r}_n=\boldsymbol{\rho}-\mathbf{d}_n$, with $\mathbf{d}_n$ the location of the
center of the $n$th particle. Here, 
regions I and II are, respectively, inside nanoparticle $1$ and $2$, and region III corresponds to
the space outside both particles (see Fig.\ \ref{fig:NPs}).

Assuming that the interparticle distance $d$ is much larger than the
nanoparticle radii $a_n$, 
we expand the Hamiltonian \eqref{eq:H_original} to $2$nd order in
$r_n/d$. Within this approximation, the expansion of the interparticle electron-electron
interaction term entering Eq.\ \eqref{eq:H_original} yields
\begin{align}
\label{eq:expansion1}
&e^2\sum_{i=1}^{N_1}\sum_{j=2}^{N_2}
\frac{1}{|\boldsymbol{\rho}_{1,i}-\boldsymbol{\rho}_{2,j}|}
\simeq
\nonumber\\
&\frac{2e^2}{d}\sum_{i=1}^{N_1}\sum_{j=1}^{N_2}
\Bigg\{
1
+\frac{(\mathbf{r}_{1,i}-\mathbf{r}_{2,j})\cdot\hat{\mathbf{d}}}{d}
\nonumber\\
&-\frac{(\mathbf{r}_{1,i}-\mathbf{r}_{2,j})\cdot\big[\mathbf{r}_{1,i}-\mathbf{r}_{2,j}
-3\hat{\mathbf{d}}\big((\mathbf{r}_{1,i}-\mathbf{r}_{2,j})\cdot\hat{\mathbf{d}}\big)\big]}{2d^2}
\Bigg\}.
\end{align}
Here, $\mathbf{r}_{n,i}$ denotes the position of the $i$th electron belonging to the
$n$th nanoparticle relative to its center. 
Similarly, the expansion of the single-particle confinement \eqref{eq:U_2NP}
yields
\begin{subequations}
\label{eq:expansion2}
\begin{align}
U_\mathrm{2NP}(\boldsymbol{\rho}_{1,i})\simeq&\; U_1(r_{1,i})-\frac{N_2e^2}{d}
\Bigg\{
1+\frac{\mathbf{r}_{1,i}\cdot\hat{\mathbf{d}}}{d}
\nonumber\\
&-\frac{\mathbf{r}_{1,i}\cdot\big[\mathbf{r}_{1,i}-3\hat{\mathbf{d}}(\mathbf{r}_{1,i}\cdot\hat{\mathbf{d}})\big]}{2d^2}
\Bigg\},
\\
U_\mathrm{2NP}(\boldsymbol{\rho}_{2,i})\simeq&\; U_2(r_{2,i})-\frac{N_1e^2}{d}
\Bigg\{
1-\frac{\mathbf{r}_{2,i}\cdot\hat{\mathbf{d}}}{d}
\nonumber\\
&-\frac{\mathbf{r}_{2,i}\cdot\big[\mathbf{r}_{2,i}-3\hat{\mathbf{d}}(\mathbf{r}_{2,i}\cdot\hat{\mathbf{d}})\big]}{2d^2}
\Bigg\},
\end{align}
\end{subequations}
where $U_n$
is the single-particle confinement of an isolated nanoparticle defined in Eq.\ \eqref{eq:U},
which is harmonic with the Mie frequency
inside the nanoparticle 
and Coulomb-like outside \cite{weick05_PRB, weick06_PRB}. Notice that
$\omega_{1}=\omega_{2}$ if the two nanoparticles are made
of the same metal, i.e., they have the same electronic density.

Using Eqs.\ \eqref{eq:expansion1} and \eqref{eq:expansion2} to express Eq.\
\eqref{eq:H_original}, we obtain the Hamiltonian 
\begin{equation}
\label{eq:H_expand}
H=\sum_{n=1}^2H_n+H_\mathrm{d\textrm{-}d}, 
\end{equation}
up to an irrelevant constant.
In Eq.\ \eqref{eq:H_expand}, 
\begin{equation}
\label{eq:H_n}
H_n=\sum_{i=1}^{N_n}\left[\frac{\mathbf{p}_{n,i}^2}{2m_\mathrm{e}}+U_n({r}_{n,i})\right]
+\frac{e^2}{2}\sum_{\substack{i,j=1\\(i\neq j)}}^{N_n}\frac{1}{|\mathbf{r}_{n,i}-\mathbf{r}_{n,j}|}
\end{equation}
represents the Hamiltonian of the isolated nanoparticle $n$ \cite{weick05_PRB, weick06_PRB}, while
\begin{equation}
\label{eq:H_d-d}
H_\mathrm{d\textrm{-}d}=\frac{e^2}{d^3}\sum_{i=1}^{N_1}\sum_{j=1}^{N_2}
\left[\mathbf{r}_{1,i}\cdot\mathbf{r}_{2,j}
-3(\mathbf{r}_{1,i}\cdot\hat{\mathbf{d}})(\mathbf{r}_{2,j}\cdot\hat{\mathbf{d}})\right]
\end{equation}
stands for the dipole-dipole interaction between the two
electron distributions in the respective nanoparticles. 
Note that retardation effects can be neglected as we assume that the
interparticle distance $d$ is much smaller than the wavelength associated with
each LSP frequency, so that the quasistatic approximation is valid.

The Hamiltonian \eqref{eq:H_expand} can be
conveniently expressed in terms of the electronic center-of-mass coordinates 
$\mathbf{R}_n=\sum_{i=1}^{N_n}\mathbf{r}_{n,i}/N_n$
and momenta
$\mathbf{P}_n=\sum_{i=1}^{N_n}\mathbf{p}_{n,i}$, 
and the relative coordinates
$\mathbf{r}'_{n,i}=\mathbf{r}_{n,i}-\mathbf{R}_n$ and 
$\mathbf{p}'_{n,i}=\mathbf{p}_{n,i}-\mathbf{P}_n/{N_n}$ ($n=1,2$) \cite{gerch02_PRA, weick06_PRB}.
Assuming that the center-of-mass displacements are much smaller than the nanoparticle radii, we obtain to second
order in the parameter $R_n/a_n\ll1$ the decomposition \eqref{eq:H_decompo}.
The center-of-mass Hamiltonian representing the plasmonic collective excitations coupled via the
dipole-dipole interaction reads
\begin{align}
\label{eq:H_cm_app}
H_\mathrm{pl}=&
\sum_{n=1}^2\left(\frac{P_n^2}{2M_n}+\frac{M_n}{2}\tilde{\omega}_{n}^2R_n^2\right)
\nonumber\\
&+\frac{Q_1Q_2}{d^3}\left[\mathbf{R}_1\cdot\mathbf{R}_2-3(\mathbf{R}_1\cdot\hat{\mathbf{d}})(\mathbf{R}_2\cdot\hat{\mathbf{d}})\right], 
\end{align}
with $M_n=N_nm_\mathrm{e}$ and $Q_n=N_ne$ the total electronic mass and charge
in the $n$th nanoparticle, respectively, and where $\tilde\omega_n$ is defined in Eq.\ \eqref{eq:omega_Mie}.
In Eq.\ \eqref{eq:H_decompo}, 
\begin{align}
\label{eq:H_rel_app}
H_\mathrm{eh}=&\sum_{n=1}^2\sum_{i=1}^{N_n}
\left[\frac{{p'_{n,i}}^2}{2m_\mathrm{e}}+U_n({r}'_{n,i})\right]
\nonumber\\
&+\frac{e^2}{2}\sum_{n=1}^2\sum_{\substack{i,j=1\\(i\neq
j)}}^{N_n}\frac{1}{|\mathbf{r}'_{n,i}-\mathbf{r}'_{n,j}|}
\end{align}
represents the Hamiltonian for the relative electronic coordinates, while
\begin{equation}
\label{eq:H_c_app}
H_\mathrm{pl-eh}=\sum_{n=1}^2\sum_{i=1}^{N_n}\mathbf{R}_n\cdot\nabla U_n(r_{n,i})\big|_{\mathbf{R}_n=0}
\end{equation}
is the coupling Hamiltonian between center-of-mass and relative coordinates.

Introducing the bosonic operator 
\begin{equation}
b_n=\frac{1}{\sqrt{2}}\left(\frac{R_n}{\ell_n}+\frac{\mathrm{i}\, P_n\ell_n}{\hbar}\right)
\end{equation}
annihilating a plasmon in nanoparticle $n$ and its adjoint 
$b_n^\dagger$, with $\ell_n=(\hbar/M_n\tilde{\omega}_n)^{1/2}$ the associated
oscillator length, the plasmonic Hamiltonian \eqref{eq:H_cm_app} transforms into Eq.\ \eqref{eq:H_pl}.
Notice that in writing the latter, we assumed that in each eigenmode the two LSPs 
are polarized in the same direction $\hat{\boldsymbol{\epsilon}}=\cos{\theta}\,
\hat{\mathbf{z}}+\sin{\theta}\, \hat{\mathbf{x}}$
forming an angle $\theta$ with the $z$ axis (see Fig.\ \ref{fig:NPs}).

Assuming that electronic correlations are not important for the present problem,
we approximate the Hamiltonian \eqref{eq:H_rel_app} by its mean-field counterpart \eqref{eq:H_env}.
Density functional
theory numerical calculations \cite{zuloa09_NL} suggest that the self-consistent potential $V$ can be approximated by two
spherical square wells of height $V_0$ centered around each nanoparticle, 
\begin{equation}
\label{eq:V_0}
V(\boldsymbol{\rho})\simeq
\begin{cases}
0, &
\boldsymbol{\rho}\in\textrm{I \& II}, 
\\
V_0, &
\boldsymbol{\rho}\in\textrm{III}.
\end{cases}
\end{equation}
Note that the form \eqref{eq:H_env} implicitly assumes that tunneling of
electrons between the two wells is suppressed. 
Within our mean-field approximation, the coupling Hamiltonian \eqref{eq:H_c_app}
thus takes the form
\eqref{eq:H_c}.
By relating collective and relative
coordinates, this expression provides the way to calculate the decay rate of the coupled plasmonic modes within a
quantum-mechanical approach.

The separation of center-of-mass and relative coordinates presented in this appendix allowed a complete quantum-mechanical treatment of the 
problem, which by essence is nonlocal. Notice that nonlocal effects in nanostructures can also be included within classical electrodynamic theories \cite{raza14_preprint}.

\section{Diagonalization of the plasmonic Hamiltonian}
\label{sec:diag}
The standard bosonic Bogoliubov transformation, applicable when the two oscillators have the same frequency, becomes more complicated 
in the case that interests us of a heterogeneous dimer. In this appendix, we follow Tsallis \cite{tsall78_JMP} for the diagonalization of the 
Hamiltonian \eqref{eq:H_pl}. Towards this purpose, we 
introduce the operators
\begin{equation}
\mathbf{b}=
\begin{pmatrix}
b_1\\
b_2\\
b_1^\dagger\\
b_2^\dagger
\end{pmatrix}, \qquad
\mathbf{b}^\dagger =
\begin{pmatrix}
b_1^\dagger & b_2^\dagger & b_1^{} & b_2^{}
\end{pmatrix}
\end{equation}
and 
\begin{equation}
\label{eq:B}
\mathbf{B}=
\begin{pmatrix}
B_+\\
B_-\\
B_+^\dagger\\
B_-^\dagger
\end{pmatrix}
=
\mathcal{T}^\dagger \mathbf{b},\quad
\mathbf{B}^\dagger=
\begin{pmatrix}
B_+^\dagger & B_-^\dagger & B_+^{} & B_-^{}
\end{pmatrix}
=\mathbf{b}^\dagger\mathcal{T}.
\end{equation}
The transformation matrix $\mathcal{T}$ is defined by
\begin{equation}
\label{eq:T}
\mathcal{T}=
\begin{pmatrix}
u_{1,+} & u_{1,-} & \bar u_{1,+} & \bar u_{1,-} \\
u_{2,+} & u_{2,-} & \bar u_{2,+} & \bar u_{2,-} \\
\bar u_{1,+} & \bar u_{1,-} & u_{1,+} & u_{1,-} \\
\bar u_{2,+} & \bar u_{2,-} & u_{2,+} & u_{2,-} 
\end{pmatrix},
\end{equation}
so that $H_\mathrm{pl}=\mathbf{b}^\dagger\mathcal{H}_\mathrm{pl}\mathbf{b}=
\mathbf{B}^\dagger \mathcal{H}_\mathrm{pl}^\mathrm{D} \mathbf{B}$ with 
\begin{equation}
\mathcal{H}_\mathrm{pl}=\frac{\hbar}{2}
\begin{pmatrix}
\tilde{\omega}_1 & \Omega f(\theta) & 0 & \Omega f(\theta) \\
\Omega f(\theta) & \tilde{\omega}_2 & \Omega f(\theta) & 0 \\
0 & \Omega f(\theta) & \tilde{\omega}_1 & \Omega f(\theta) \\
\Omega f(\theta) & 0 & \Omega f(\theta) & \tilde{\omega}_2
\end{pmatrix}
\end{equation}
and 
\begin{equation}
\label{eq:H_diag}
\mathcal{H}_\mathrm{pl}^\mathrm{D}=\frac{\hbar}{2}
\begin{pmatrix}
\omega_+ & 0 & 0 & 0 \\
0 & \omega_- & 0 & 0 \\
0 & 0 & \omega_+ & 0 \\
0 & 0 & 0 & \omega_-
\end{pmatrix}, 
\end{equation}
up to irrelevant constants. Imposing that the new operators $B_\pm$ of Eq.\ \eqref{eq:B} are
bosonic, the coefficients entering the transformation matrix $\mathcal{T}$ defined in Eq.\
\eqref{eq:T} obey
\begin{equation}
\label{eq:condition}
\sum_{n=1}^2 \left(u_{n,\pm}^2-\bar u_{n,\pm}^2\right)=1.
\end{equation}

Obtaining the
diagonal form \eqref{eq:H_diag} amounts to diagonalizing the matrix
$2\mathcal{H}_\mathrm{pl}\mathcal{J}$, where 
\begin{equation}
\mathcal{J}=
\begin{pmatrix}
1 & 0 & 0 & 0 \\
0 & 1 & 0 & 0 \\
0 & 0 & -1 & 0 \\
0 & 0 & 0 & -1 
\end{pmatrix}.
\end{equation}
The condition $\det{\{2\mathcal{H}_\mathrm{pl}\mathcal{J}-\hbar\omega\mathbb{1}\}}=0$ yields the
eigenvalue equation 
\begin{equation}
\label{eq:eigenvalue}
\left(\omega^2-\tilde{\omega}_1^2\right)\left(\omega^2-\tilde{\omega}_2^2\right)
=4\Omega^2\tilde{\omega}_1\tilde{\omega}_2f^2(\theta).
\end{equation}
Solving for $\omega$, we find the eigenfrequencies \eqref{eq:omega_pm}
of the coupled plasmonic modes. The eigenvectors of
$2\mathcal{H}_\mathrm{pl}\mathcal{J}$ then 
determine the coefficients of the transformation matrix \eqref{eq:T} through
\begin{equation}
\label{eq:system}
\begin{pmatrix}
\tilde{\omega}_1-\omega_\pm & \Omega f(\theta) & 0 & \Omega f(\theta) \\
\Omega f(\theta) & \tilde{\omega}_2-\omega_\pm & \Omega f(\theta) & 0 \\
0 & \Omega f(\theta) & -\tilde{\omega}_1-\omega_\pm & -\Omega f(\theta) \\
\Omega f(\theta) & 0 & -\Omega f(\theta) & -\tilde{\omega}_2-\omega_\pm
\end{pmatrix}
\hspace{-.2truecm}
\begin{pmatrix}
u_{1,\pm} \\
u_{2,\pm} \\
\bar u_{1,\pm}\\
\bar u_{2,\pm}\\
\end{pmatrix}
=0.
\end{equation}
Solving for the system \eqref{eq:system}, we find together with the condition
\eqref{eq:condition} and with the help of the eigenvalue equation
\eqref{eq:eigenvalue} the coefficients \eqref{eq:coeff}
entering the transformation matrix \eqref{eq:T}.



\begin{thebibliography}{}
\bibitem{footnote:d}
This dependence on $d$ strictly holds in the regime
$|\tilde\omega_1^2-\tilde\omega_2^2|/2\tilde\omega_1\tilde\omega_2\ll2\Omega|f(\theta)|/(\tilde\omega_1\tilde\omega_2)^{1/2}\ll1$.
In the case
$2\Omega|f(\theta)|/(\tilde\omega_1\tilde\omega_2)^{1/2}\ll|\tilde\omega_1^2-\tilde\omega_2^2|/2\tilde\omega_1\tilde\omega_2\ll1$,
we have $\omega_+\simeq\max{\{\tilde\omega_1,\tilde\omega_2\}}$ and
$\omega_-\simeq\min{\{\tilde\omega_1,\tilde\omega_2\}}$.
\bibitem{footnote:theta_0}
Notice that even in the case where
the dipole-dipole interaction vanishes ($\theta=\theta_0$),
$\gamma_\pm^\mathrm{L}$ is not simply equal to the linewidth $\gamma_n^\mathrm{L}$ of an isolated
nanoparticle defined in Eq.\ \eqref{eq:kubo}. This effect is due to the modification of the non-harmonic part of the single-particle
confinement \eqref{eq:U} by the other nanoparticle  and the
subsequent breaking of Kohn's theorem.
\bibitem{footnote:noble}
The Mie frequency defined in Eq.\ \eqref{eq:omega_Mie}, depending on the dielectric constants inside
and outside the nanoparticle, provides in general a good approximation to the experimentally obtained values
(though deviations may appear in the presence of interband transitions close to the resonance
frequency). The precise value of the Landau damping rate in noble metals is
difficult to determine since in the case of
very small nanoparticles, it depends on the softness of the self-consistent
confining potential. This is due
to the dielectric mismatch at the nanoparticle boundary, see Ref.\ \cite{weick06_PRB}.
\end{thebibliography}
\end{document}